\batchmode
\makeatletter
\def\input@path{{"C:/Trabajo laptop/Mis articulos/Fusion 2026/Variational PMB filter/"}}
\makeatother
\documentclass[onecolumn,english,twocolumn]{IEEEtran}
\usepackage[T1]{fontenc}
\usepackage[latin9]{inputenc}
\usepackage{color}
\usepackage{float}
\usepackage{amsmath}
\usepackage{amsthm}
\usepackage{amssymb}
\usepackage{graphicx}
\PassOptionsToPackage{normalem}{ulem}
\usepackage{ulem}

\makeatletter

\providecommand{\tabularnewline}{\\}
\floatstyle{ruled}
\newfloat{algorithm}{tbp}{loa}
\providecommand{\algorithmname}{Algorithm}
\floatname{algorithm}{\protect\algorithmname}

\theoremstyle{plain}
\newtheorem{thm}{\protect\theoremname}
\theoremstyle{definition}
\newtheorem{defn}[thm]{\protect\definitionname}
\theoremstyle{plain}
\newtheorem{lem}[thm]{\protect\lemmaname}

\usepackage{cite} 
\usepackage[margin=8pt,font=footnotesize]{caption}
\usepackage{algorithm}
\usepackage{algpseudocode}

\usepackage{amsmath}  
\allowdisplaybreaks

\makeatother

\usepackage{babel}
\providecommand{\definitionname}{Definition}
\providecommand{\lemmaname}{Lemma}
\providecommand{\theoremname}{Theorem}

\begin{document}
\title{Variational PMB filter via coordinate descent Kullback-Leibler divergence
minimisation}
\author{Ángel F. García-Fernández$^{\star}$, Yuxuan Xia$^{\circ}$\\
{\normalsize $^{\star}$Information Processing and Telecommunications
Center, Universidad Politécnica de Madrid, Madrid, Spain}\\
{\normalsize $^{\circ}$School of Automation and Intelligent Sensing,
Shanghai Jiao Tong University, Shanghai, China}\\
{\normalsize Emails: angel.garcia.fernandez@upm.es, yuxuan.xia@sjtu.edu.cn}\thanks{This work was supported by the Spanish Ministry of Science, Innovation and Universities under the project PID2024-158149OB-C21.}}

\maketitle
\thispagestyle{empty}
\begin{abstract}
This paper presents a new derivation of the variational Poisson multi-Bernoulli
(V-PMB) filter for multi-target estimation proposed in \cite{Williams15}.
The proposed derivation is based on considering an augmented space
that includes the set of target states with their track indices and
the global hypothesis variable. Then, we show that the V-PMB projection
performs a coordinate descent Kullback-Leibler divergence (KLD) minimisation
on this augmented space to fit the best possible PMB density to the
Poisson multi-Bernoulli mixture (PMBM) posterior. We also show that
this V-PMB projection keeps the probability hypothesis density of
the posterior. The paper also includes a comparison with the PMBM
filter and other PMB filter variants, including a track-oriented Murty-based
implementation, a track-oriented loopy belief propagation implementation
and a global nearest neighbour implementation, showing the benefits
of the V-PMB filter compared to the other PMB filters when targets
get in close proximity and then separate.
\end{abstract}


\section{Introduction}

Multi-target filtering refers to the estimation of the states of multiple
dynamic targets given noisy sensor data. It has application in multiple
domains including autonomous vehicles \cite{Scheidegger18}, maritime
traffic monitoring \cite{Brekke21} and space situational awareness
\cite{Delande19}. In this paper, we consider a Bayesian approach
for multi-target filtering in which there are probabilistic models
for target births, dynamics and deaths, and also for the sensor measurements
\cite{Mahler_book14}. Then, all information about the current set
of targets is included in its multi-target density given all past
measurements (the posterior density). 

When the sensor data consists of detections, there are two main measurement
models: point-target measurement model, when each target generates
at most one detection, and extended-target measurement model, when
each target can generate multiple measurements \cite{Mahler_book14}.
Both measurement models assume that the clutter process is a Poisson
point process (PPP). If we consider the standard multi-target dynamic
model with PPP birth \cite{Mahler_book14}, the posterior density
of the set of targets for both measurement models is a Poisson multi-Bernoulli
mixture (PMBM) \cite{Williams15b,Angel18_b,Granstrom20}. A PMBM density
represents the union of two independent processes, the PPP process
and a multi-Bernoulli mixture (MBM) process. The PPP contains information
on targets that remain undetected. Each mixture component in the MBM
represents the information on the current targets given a global data
association hypothesis. The posterior is also a PMBM for general target-generated
measurement models and arbitrary clutter distributions \cite{Angel23}.

While PMBM filtering theoretically provides the posterior density,
its computation depends on the number of data associations, which
rapidly becomes very large. This implies that approximations such
as gating, pruning global hypotheses and pruning Bernoulli components
are necessary in practice \cite{Angel18_b}. An alternative is to
design Poisson multi-Bernoulli (PMB) filters that only consider one
term in the MBM. One option is the global nearest neighbour (GNN)
PMB filter, which selects the most likely data association at each
update step. This results in a fast implementation though it can have
difficulties dealing with challenging data association problems, as
in a high cluttered scenario. Another alternative is the (track-oriented)
PMB filter \cite{Williams15b}. The PMB filter can be derived as a
minimisation of the Kullback-Leibler divergence (KLD) between the
updated PMBM and the considered PMB, once the target states have been
augmented with track indices, which are auxiliary variables in the
PMBM posterior \cite{Angel20_e}. For fixed and known number of targets,
the Gaussian implementation of the PMB filter results in the joint
probabilistic data association (JPDA) filter \cite{Fortmann83}. In
addition, the PMB filter is a fully Bayesian variant of the joint
integrated probabilistic data association (JIPDA) filters \cite{Musicki04}
for an unknown and variable number of targets \cite{Williams15b,Brekke21}.
The PMB filter can be implemented, for instance, by obtaining the
global hypotheses with highest weights (e.g., using Murty's algorithm
\cite{Murty68}) or by using (loopy) belief propagation \cite{Williams14,Meyer18,Kim24}.
A drawback of the PMB filter is that its Gaussian implementation can
suffer from track coalescence when there are closely spaced targets.

To improve performance in situations when there are closely spaced
targets, several related multi-target filtering algorithms have been
proposed. For fixed and known number of targets, the Gaussian approximation
in the JPDA filter can be improved by noticing that the information
on the set of targets is invariant under permutation \cite{Svensson11}.
Taking advantage of this property and using a KLD cost function led
to development the set JPDA algorithm \cite{Svensson11}. The permutation-invariant
property was also used for improving multi-Bernoulli approximations
using particle filters for track-before-detect in \cite{Angel16}. 

References \cite{Williams15,Williams14b} introduced variational PMB
filters for point-target measurement models that can improve the (track-oriented)
PMB filter when targets get in close proximity. The variational PMB
filter projects the updated PMBM into a PMB density by first introducing
a missing distribution on the permutation of the Bernoulli components
for each global hypothesis. Then, an upper bound of the KLD between
the posterior PMBM and the looked-for PMB approximation is obtained.
This KLD upper bound depends on the missing distribution and a scalar
parameter, called temperature $T\in[0,1]$. Of particular interest
to this paper is the zero temperature case \cite[Sec. III.C]{Williams15}.
In this case, the optimal missing permutation distribution takes a
form of winner-takes all, and is equivalent to finding the best permutation
for each global hypothesis. We refer to the PMB filter with this update-projection
step as the variational PMB (V-PMB) filter. The V-PMB projection also
admits an approximate, computationally lighter implementation \cite[Sec. III.D]{Williams15}.
The V-PMB filter has been shown to deal well with targets in close
proximity and has also been applied to extended targets \cite[Sec. V. B]{Xia22}. 

\begin{figure}
\begin{centering}
\includegraphics[scale=0.9]{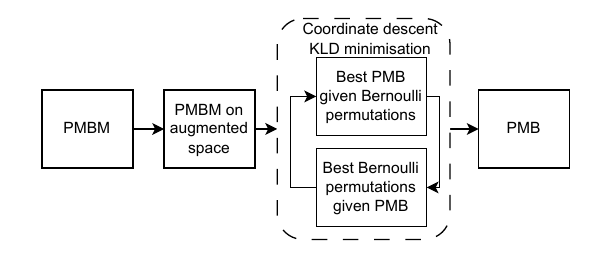}
\par\end{centering}
\caption{\label{fig:Diagram-VPMB_projection}Diagram of the V-PMB projection
of a PMBM density via coordinate descent KLD minimisation.}

\end{figure}

The contribution of this paper is a new derivation of the V-PMB projection
step with the objective of providing a better understanding of this
important algorithm and of the use of auxiliary variables in multi-target
inference. To do so, we start from the PMBM posterior, which is then
written on an augmented space consisting of the set of targets, with
their track indices, and the global hypothesis variable \cite{Xia23}.
Then, the V-PMB projection step is derived as a coordinate descent
KLD minimisation by optimising over the best PMB approximation and
the best permutations iteratively, as represented in the diagram shown
in Figure \ref{fig:Diagram-VPMB_projection}.  We also show that
the V-PMB projection step keeps the probability hypothesis density
(PHD), also called first-moment density \cite{Mahler_book14}, of
the PMBM distribution. Finally, we provide simulations demonstrating
the advantages of the V-PMB filter compared to other PMB filters.

\section{Problem formulation}

In this section, we first present the PMBM density on the set of targets
in Section \ref{subsec:PMBM-density}. Then, we consider the augmented
space with the set of targets with track indices and the global hypothesis
variable, and write the corresponding PMBM density in Section \ref{subsec:PMBM-density-auxiliary}.
 Finally, we present the problem formulation as the calculation of
the best fitting PMB density in the augmented space that minimises
the KLD in Section \ref{subsec:Best-PMB-approximation}.

\subsection{PMBM density\label{subsec:PMBM-density}}

A target state is denoted by $x\in\mathbb{R}^{n_{x}}$. The set of
targets at time step $k$ is denoted by $X_{k}=\left\{ x_{k}^{1},...,x_{k}^{|X_{k}|}\right\} $
such that $X_{k}\in\mathcal{F}\left(\mathbb{R}^{n_{x}}\right)$, which
denotes the set of all finite subsets of $\mathbb{R}^{n_{x}}$ \cite{Mahler_book14}.
The measurement model can be any detection-based model that results
in a PMBM posterior \cite{Angel23}. After processing the sequence
of measurements up to time step $k'\in\{k-1,k\}$, the density of
the set of targets at time step $k$ is a PMBM of the form
\begin{align}
f_{k|k'}\left(X_{k}\right) & =\sum_{Y\uplus W=X_{k}}f_{k|k'}^{\mathrm{p}}\left(Y\right)f_{k|k'}^{\mathrm{mbm}}\left(W\right),\label{eq:PMBM}\\
f_{k|k'}^{\mathrm{p}}\left(X_{k}\right) & =e^{-\int\lambda_{k|k'}\left(x\right)dx}\prod_{x\in X_{k}}\lambda_{k|k'}\left(x\right),\\
f_{k|k'}^{\mathrm{mbm}}\left(X_{k}\right) & =\sum_{a\in\mathcal{A}_{k|k'}}w_{k|k'}^{a}\sum_{\uplus_{l=1}^{n_{k|k'}}X^{l}=X_{k}}\prod_{i=1}^{n_{k|k'}}f_{k|k'}^{i,a^{i}}\left(X^{i}\right),\label{eq:MBM}
\end{align}
where the sum in (\ref{eq:PMBM}) is a convolution sum that goes through
all disjoint and possibly empty sets $Y$ and $W$ such that their
union is $X_{k}$. The PPP density is $f_{k|k'}^{\mathrm{p}}\left(\cdot\right)$
with intensity $\lambda_{k|k'}\left(\cdot\right)$. The number of
potential targets is $n_{k|k'}$. The $i$-th potential target has
local hypotheses with indices taking values in $\left\{ 1,...,h_{k|k'}^{i}\right\} $,
where $h_{k|k'}^{i}$ is the number of local hypotheses. The Bernoulli
density of the $i$-th potential target given its local hypothesis
$a^{i}\in\left\{ 1,...,h_{k|k'}^{i}\right\} $ is $f_{k|k'}^{i,a^{i}}\left(\cdot\right)$
and has a probability of existence $r_{k|k'}^{i,a^{i}}$ and single-target
density $p_{k|k'}^{i,a^{i}}\left(\cdot\right)$. The MBM density $f_{k|k'}^{\mathrm{mbm}}\left(\cdot\right)$
goes through all global hypotheses $a=(a^{1},...,a^{n_{k|k'}})$ belonging
to the set of all global hypotheses $\mathcal{A}_{k|k'}$. The weight
of the global hypothesis $a$ is $w_{k|k'}^{a}$. Full details on
how these parameters are propagated through the filtering recursion
and the definition of global hypotheses depending on the measurement
model are provided in previous works \cite{Williams15b,Angel18_b,Granstrom20,Angel23}.

To derive the V-PMB projection, we introduce a permutation $\pi_{a}=\left(\pi_{a}(1),...,\pi_{a}(n_{k|k'})\right)$
of the index set $\left\{ 1,...,n_{k|k'}\right\} $ for each global
hypothesis $a$, and write the MBM density as
\begin{align}
f_{k|k'}^{\mathrm{mbm}}\left(X_{k}\right) & =\sum_{a\in\mathcal{A}_{k|k'}}w_{k|k'}^{a}\sum_{\uplus_{l=1}^{n_{k|k'}}X^{l}=X_{k}}\prod_{i=1}^{n_{k|k'}}f_{k|k'}^{\pi_{a}(i),a^{\pi_{a}(i)}}\left(X^{i}\right).\label{eq:MBM_permutation}
\end{align}
That is, the PMBM density is invariant to the permutation of the Bernoulli
components in each global hypothesis.

\subsection{PMBM density with auxiliary variables\label{subsec:PMBM-density-auxiliary}}

First, we introduce some additional variables. We augment the single
target space with an auxiliary variable $u\in\mathbb{\mathbb{U}}_{k|k'}=\left\{ 0,1,..,n_{k|k'}\right\} $,
such that $\left(u,x\right)\in\mathbb{\mathbb{U}}_{k|k'}\times\mathbb{R}^{n_{x}}$
\cite{Angel20_e}. The auxiliary variable $u$ is a track index: for
$u=i\in\left\{ 1,..,n_{k|k'}\right\} $, $x$ represents the state
of the $i$-th potential target (also referred to as track \cite{Williams15b})
and, for $u=0$, $x$ represents an undetected target. A set of targets
with track indices is then $\widetilde{X}_{k}=\left\{ \left(u_{k}^{1},x_{k}^{1}\right),...,\left(u_{k}^{|X_{k}|},x_{k}^{|X_{k}|}\right)\right\} \in\mathcal{F}\left(\mathbb{\mathbb{U}}_{k|k'}\times\mathbb{R}^{n_{x}}\right)$.
 Then, the augmented variable containing the set of targets with
track indices and global hypothesis variable is $\left(\widetilde{X}_{k},a\right)\in\mathcal{F}\left(\mathbb{\mathbb{U}}_{k|k'}\times\mathbb{R}^{n_{x}}\right)\times\mathcal{A}_{k|k'}$.
Let $\pi$ be a sequence that contains the permutations for all global
hypotheses, $\pi=\left(\pi_{a}\right)_{a\in\mathcal{A}_{k|k'}}$. 
\begin{defn}
Given a PMBM density $f_{k|k'}\left(\cdot\right)$ of the form (\ref{eq:PMBM}),
with MBM density in (\ref{eq:MBM_permutation}), and a permutation
$\pi_{a}$ for each global hypothesis $a$, we define the density
$\widetilde{f}_{k|k'}^{\pi}\left(\cdot\right)$ on the augmented variable
$\left(\widetilde{X}_{k},a\right)$ as \cite{Xia23}
\begin{align}
 & \widetilde{f}_{k|k'}^{\pi}\left(\widetilde{X}_{k},a\right)\nonumber \\
 & =\sum_{\uplus_{l=1}^{n_{k|k'}}\widetilde{X}^{l}\uplus\widetilde{Y}=\widetilde{X}_{k}}\widetilde{f}_{k|k'}^{p}\left(\widetilde{Y}\right)w_{k|k'}^{a}\prod_{i=1}^{n_{k|k'}}\left[\widetilde{f}_{k|k'}^{\pi_{a}(i),a^{\pi_{a}(i)}}\left(\widetilde{X}^{i}\right)\right]\\
 & =\widetilde{f}_{k|k'}^{p}\left(\widetilde{Y}_{k}\right)w_{k|k'}^{a}\prod_{i=1}^{n_{k|k'}}\left[\widetilde{f}_{k|k'}^{\pi_{a}(i),a^{\pi_{a}(i)}}\left(\widetilde{X}_{k}^{i}\right)\right]\label{eq:PMBM_aux_var2}
\end{align}
where, for a given $\widetilde{X}_{k}$, $\widetilde{Y}_{k}=\left\{ \left(u,x\right)\in\widetilde{X}_{k}:u=0\right\} $
and $\widetilde{X}_{k}^{i}=\left\{ \left(u,x\right)\in\widetilde{X}_{k}:u=i\right\} $,
and
\begin{align}
\widetilde{f}_{k|k'}^{p}\left(\widetilde{Y}_{k}\right) & =e^{-\int\lambda_{k|k'}\left(x\right)dx}\left[\widetilde{\lambda}_{k|k'}\left(\cdot\right)\right]^{\widetilde{Y}_{k}}\label{eq:PPP_augmented}\\
\widetilde{\lambda}_{k|k'}\left(u,x\right) & =\delta_{0}\left[u\right]\lambda_{k|k'}\left(x\right)
\end{align}
\begin{align}
 & \widetilde{f}_{k|k'}^{\pi_{a}(i),a^{\pi_{a}(i)}}\left(\widetilde{X}_{k}^{i}\right)\nonumber \\
 & \quad=\begin{cases}
1-r_{k|k'}^{\pi_{a}(i),a^{\pi_{a}(i)}} & \widetilde{X}_{k}^{i}=\emptyset\\
r_{k|k'}^{\pi_{a}(i),a^{\pi_{a}(i)}}p_{k|k'}^{\pi_{a}(i),a^{\pi_{a}(i)}}\left(x\right)\delta_{i}\left[u\right] & \widetilde{X}_{k}^{i}=\left\{ \left(u,x\right)\right\} \\
0 & \mathrm{otherwise}
\end{cases}\label{eq:Bernoulli_density_filter-aux}
\end{align}
where the Kronecker delta $\delta_{i}\left[u\right]=1$ if $u=i$
and $\delta_{i}\left[u\right]=0$, otherwise.
\end{defn}
It should be noted that, if we choose $\pi_{a}=\left(1,...,n_{k|k'}\right)$
for all $a\in\mathcal{A}_{k|k'}$ and sum over $a\in\mathcal{A}_{k|k'}$,
we recover the definition of the PMBM with auxiliary variables in
\cite{Angel20_e}. If we sum over $a\in\mathcal{A}_{k|k'}$ and integrate
out the track index $u$ for each target, following \cite{Angel20_e},
we recover the original PMBM density (\ref{eq:PMBM}), as required
when auxiliary variables are introduced in a density \cite{Pitt99}.
In addition, in $\widetilde{X}_{k}$, there can be multiple targets
with $u=0$, but at most one target with $i\in\left\{ 1,..,n_{k|k'}\right\} $,
to yield a non-zero density $\widetilde{f}_{k|k'}^{\pi}\left(\widetilde{X}_{k},a\right)$.

\subsection{Best PMB approximation via KLD minimisation\label{subsec:Best-PMB-approximation}}

We aim to obtain a PMB approximation $\widetilde{q}\left(\cdot\right)$
on the augmented space $\mathcal{F}\left(\mathbb{\mathbb{U}}_{k'|k}\times\mathbb{R}^{n_{x}}\right)\times\mathcal{A}_{k|k'}$
such that
\begin{align}
\widetilde{q}\left(\widetilde{X}_{k},a\right) & =w_{q}^{a}\widetilde{q}^{p}\left(\widetilde{Y}_{k}\right)\prod_{i=1}^{n_{k|k'}}\left[\widetilde{q}^{i}\left(\widetilde{X}_{k}^{i}\right)\right]\label{eq:PMB_approx}
\end{align}
where $\widetilde{q}^{p}\left(\cdot\right)$ is the PPP density, $\widetilde{q}^{i}\left(\cdot\right)$
is the $i$-th Bernoulli density, and $w_{q}^{a}$ is the weight of
variable $a$ such that 
\begin{align}
\sum_{a\in\mathcal{A}_{k|k'}}w_{q}^{a} & =1.
\end{align}
 Importantly, the PPP and Bernoulli components do not depend on
$a$. This implies that, as required, by integrating out the global
hypothesis $a$, we obtain the PMB density with auxiliary variables
in \cite{Angel20_e}. If we additionally integrate out the track index
in each single-target state, we obtain a standard PMB density \cite{Angel20_e}. 

Our objective is to minimise the KLD
\begin{align}
\left(\widetilde{q}^{*},\pi^{*}\right) & =\underset{\left(\widetilde{q},\pi\right)}{\mathrm{argmin}}\,\mathrm{D}\left(\widetilde{f}_{k|k'}^{\pi}\left\Vert \widetilde{q}\right.\right)\label{eq:KLD_optimisation_problem}
\end{align}
where
\begin{align}
 & \mathrm{D}\left(\widetilde{f}_{k|k'}^{\pi}\left\Vert \widetilde{q}\right.\right)\nonumber \\
 & \quad=\sum_{a\in\mathcal{A}_{k|k'}}\int\widetilde{f}_{k|k'}^{\pi}\left(\widetilde{X}_{k},a\right)\log\frac{\widetilde{f}_{k|k'}^{\pi}\left(\widetilde{X}_{k},a\right)}{\widetilde{q}\left(\widetilde{X}_{k},a\right)}\delta\widetilde{X}_{k}.\label{eq:KLD_augmented_space}
\end{align}

\section{PMB projection via coordinate descent KLD minimisation\label{sec:PMB-projection-KLD_minimisation}}

In this section, we develop a coordinate descent algorithm to iteratively
minimise (\ref{eq:KLD_optimisation_problem}). We first provide a
simplification of the KLD in Section \ref{subsec:KLD-simplification}.
Then, Sections \ref{subsec:Optimisation-over-q} and \ref{subsec:Optimisation-permutations}
present the optimisations over $\widetilde{q}\left(\cdot\right)$
and over the permutations, respectively. The PHD matching property
is provided in Section \ref{subsec:Matching-the-PHD}. Section \ref{subsec:Discussion}
discusses the derivation. 

\textcolor{red}{}

\subsection{KLD simplification\label{subsec:KLD-simplification}}

We present the following lemma that provides a neat expression of
the KLD (\ref{eq:KLD_augmented_space}).
\begin{lem}
\label{lem:KLD_simplified}Let $\widetilde{f}_{k|k'}^{\pi}\left(\cdot\right)$
be a PMBM density of the form (\ref{eq:PMBM_aux_var2}) and $\widetilde{q}\left(\cdot\right)$
be a PMB density of the form (\ref{eq:PMB_approx}), both defined
on the augmented space $\mathcal{F}\left(\mathbb{\mathbb{U}}_{k|k'}\times\mathbb{R}^{n_{x}}\right)\times\mathcal{A}_{k|k'}$.
The KLD $\mathrm{D}\left(\widetilde{f}_{k|k'}^{\pi}\left\Vert \widetilde{q}\right.\right)$
in (\ref{eq:KLD_augmented_space}) can be written as
\begin{align}
\mathrm{D}\left(\widetilde{f}_{k|k'}^{\pi}\left\Vert \widetilde{q}\right.\right) & =\mathrm{D}\left(\widetilde{f}_{k|k'}^{p}\left\Vert \widetilde{q}^{p}\right.\right)+\mathrm{D}\left(w_{k|k'}^{a}\left\Vert w_{q}^{a}\right.\right)\nonumber \\
 & \quad+\sum_{a\in\mathcal{A}_{k|k'}}w_{k|k'}^{a}\sum_{i=1}^{n_{k|k'}}\mathrm{D}\left(\widetilde{f}_{k|k'}^{\pi_{a}(i),a^{\pi_{a}(i)}}\left\Vert \widetilde{q}^{i}\right.\right).\label{eq:KLD_simplified}
\end{align}
\end{lem}
Lemma \ref{lem:KLD_simplified} is proved in Appendix \ref{sec:AppendixA}.
We can see that the KLD has three parts: KLD between PPP components,
KLD between global hypothesis weights, and sum over all global hypothesis
each with its weight and weighted sum over the KLD between Bernoulli
components \cite{Fontana23}, arranged according to permutation $\pi_{a}$.
The minimum $w_{q}^{a}$ is achieved by setting it to $w_{k|k'}^{a}$.
However, the choice of $w_{q}^{a}$ does not affect the resulting
PMB approximation.

\subsection{Optimisation over $\widetilde{q}$\label{subsec:Optimisation-over-q}}

 We start the iterated optimisation procedure with an initial permutation
$\pi_{a}^{(0)}$ for all $a\in\mathcal{A}_{k|k'}$, for instance,
with $\pi_{a}^{(0)}=\left(1,...,n_{k|k'}\right)$. The following lemma
indicates how to obtain the optimal PMB approximation at iteration
$j$.
\begin{lem}
\label{lem:Optimisation_q}Given the permutations $\pi^{(j)}$ for
all global hypotheses, the optimal PMB approximation $\widetilde{q}^{(j)}\left(\cdot\right)$
of the form (\ref{eq:PMB_approx}) at iteration $j$, 
\begin{align}
\widetilde{q}^{(j)} & =\underset{\widetilde{q}}{\mathrm{argmin}}\,\mathrm{D}\left(\widetilde{f}_{k|k'}^{\pi^{(j)}}\left\Vert \widetilde{q}\right.\right),\label{eq:optimisation_q}
\end{align}
has PPP density $\widetilde{q}^{p}\left(\cdot\right)=\widetilde{f}_{k|k'}^{p}\left(\cdot\right)$
and the $i$-th Bernoulli density is
\begin{align}
\widetilde{q}^{(j),i}\left(\widetilde{X}_{k}^{i}\right) & =\sum_{a\in\mathcal{A}_{k|k'}}w_{k|k'}^{a}\left[\widetilde{f}_{k|k'}^{\pi_{a}^{(j)}(i),a^{\pi_{a}^{(j)}(i)}}\left(\widetilde{X}_{k}^{i}\right)\right]
\end{align}
resulting in the parameters
\begin{align}
\lambda^{q}\left(x\right) & =\lambda_{k|k'}\left(x\right)\label{eq:lambda_q}\\
r^{(j),i} & =\sum_{a\in\mathcal{A}_{k|k'}}w_{k|k'}^{a}r_{k|k'}^{\pi_{a}^{(j)}(i),a^{\pi_{a}^{(j)}(i)}}\label{eq:prob_existence_q}\\
p^{(j),i}\left(x\right) & =\frac{1}{r^{(j),i}}\sum_{a\in\mathcal{A}_{k|k'}}w_{k|k'}^{a}r_{k|k'}^{\pi_{a}^{(j)}(i),a^{\pi_{a}^{(j)}(i)}}p_{k|k'}^{\pi_{a}^{j}(i),a^{\pi_{a}^{(j)}(i)}}\left(x\right).\label{eq:single_target_density_q}
\end{align}
\end{lem}
Lemma \ref{lem:Optimisation_q} is proved in Appendix \ref{sec:AppendixB}.
It should be noted that in the case that $\pi_{a}^{(j)}=\left(1,...,n_{k|k'}\right)$
for all $a\in\mathcal{A}_{k|k'}$, as done at iteration 0 in the implementation
we use in the simulations, we can simplify (\ref{eq:prob_existence_q})
and (\ref{eq:single_target_density_q}) by summing over the local
hypotheses of each Bernoulli, as done in the standard (track-oriented)
PMB approximation \cite[Prop. 1]{Angel20_e,Williams15b}. 

Integrating out $a$ and the track indices in $\widetilde{q}^{(j)}\left(\cdot\right)$
yields a PMB density
\begin{align}
q^{(j)}\left(X_{k}\right) & =\sum_{\uplus_{l=0}^{n_{k|k'}}X^{l}=X_{k}}f_{k|k'}^{\mathrm{p}}\left(X^{0}\right)\prod_{i=1}^{^{n_{k|k'}}}q^{(j),i}\left(X^{i}\right)\label{eq:PMB_iteration_j}
\end{align}
where the probability of existence and single-target density of the
Bernoulli density $q^{(j),i}\left(\cdot\right)$ are (\ref{eq:prob_existence_q})
and (\ref{eq:single_target_density_q}), respectively.

\subsection{Optimisation over the permutations\label{subsec:Optimisation-permutations}}

The following lemma indicates how to obtain the optimal permutations
at iteration $j+1$.
\begin{lem}
\label{lem:Optimisation_pi}Given the PMB approximation $\widetilde{q}^{(j)}\left(\cdot\right)$
at iteration $j$, the optimal sequence of permutations $\pi^{(j+1)}=\left(\pi_{a}^{(j+1)}\right)_{a\in\mathcal{A}_{k|k'}}$
at iteration $j+1$,
\begin{align}
\pi^{(j+1)} & =\underset{\pi}{\mathrm{argmin}}\,\mathrm{D}\left(\widetilde{f}_{k|k'}^{\pi}\left\Vert \widetilde{q}^{(j)}\right.\right)
\end{align}
can be obtained by optimising the permutation for each global hypothesis
independently of the rest by obtaining
\begin{align}
\pi_{a}^{(j+1)} & =\underset{\pi_{a}}{\mathrm{argmin}}\,\sum_{i=1}^{n_{k|k'}}\mathrm{D}\left(\widetilde{f}_{k|k'}^{\pi_{a}(i),a^{\pi_{a}(i)}}\left\Vert \widetilde{q}^{(j),i}\right.\right).\label{eq:optimisation_permutation}
\end{align}
\end{lem}
Lemma \ref{lem:Optimisation_pi} is a direct result of the KLD form
in (\ref{eq:KLD_simplified}), which is additive over $\pi_{a}$,
so we can optimise each global hypothesis independently of the rest.
It should be noted that, for each global hypothesis Lemma \ref{lem:Optimisation_pi}
selects the permutation $\pi_{a}$ that minimises the sum of the KLDs
between Bernoulli densities \cite{Fontana23} in $\widetilde{f}_{k|k'}^{\pi}\left(\cdot\right)$
and $\widetilde{q}^{(j)}\left(\cdot\right)$. Problem (\ref{eq:optimisation_permutation})
is a 2-D assignment problem for which well-established solvers exist
\cite{Crouse16}. We can now iterate over the optimisation problems
(\ref{eq:optimisation_q}) and (\ref{eq:optimisation_permutation})
until convergence. 

\subsection{Convergence criterion\label{subsec:Convergence-criterion}}

In this section, we explain the convergence criterion to determine
when to stop the iterated optimisations. In the overall KLD in (\ref{eq:KLD_simplified}),
the first two terms are constants that do not change with optimisation
iteration $j$. Therefore, we can stop the iterations if the decrease
in the third term
\begin{align}
c^{(j)} & =\sum_{a\in\mathcal{A}_{k|k'}}w_{k|k'}^{a}\sum_{i=1}^{n_{k|k'}}\mathrm{D}\left(f_{k|k'}^{\pi_{a}(i),a^{\pi_{a}(i)}}\left\Vert q^{(j),i}\right.\right)\label{eq:convergence_criterion}
\end{align}
is smaller than a predefined threshold $\Gamma$. That is, the algorithm
stops at iteration $j$ if 
\begin{align}
\left|c^{(j)}-c^{(j-1)}\right| & \leq\Gamma.\label{eq:convergence_criterion2}
\end{align}

Note that, if the permutations remain unchanged from iteration $j-1$
to $j$, then $c^{(j)}=c^{(j-1)}$, and the algorithm has converged.
In addition, the quality of the PMB approximation improves (or remains
unchanged) with each iteration, as measured by the KLD (\ref{eq:KLD_simplified}).

\subsection{Matching the PHD\label{subsec:Matching-the-PHD}}

In this section, we show that the coordinate descent algorithm does
not change the PHD of the PMBM (\ref{eq:PMBM}). The PHD of the PMBM
is the sum of the PHD of the PPP and the MBM resulting in \cite{Mahler_book14}
\begin{align}
\mathrm{D}_{f_{k|k'}}\left(x\right) & =\lambda_{k|k'}\left(x\right)+\sum_{a\in\mathcal{A}_{k|k'}}w_{k|k'}^{a}\nonumber \\
 & \quad\times\sum_{i=1}^{n_{k|k'}}r_{k|k'}^{\pi_{a}(i),a^{\pi_{a}(i)}}p_{k|k'}^{\pi_{a}(i),a^{\pi_{a}(i)}}\left(x\right).\label{eq:PMBM_PHD}
\end{align}

The PHD of the PMB at iteration $j$, given by (\ref{eq:PMB_iteration_j}),
is given by the sum of the PHD of the PPP and the PHDs of all Bernoulli
components, which results in \cite{Mahler_book14}
\begin{align}
\mathrm{D}_{q^{(j)}}\left(x\right) & =\lambda_{k|k'}\left(x\right)+\sum_{i=1}^{n_{k|k'}}\sum_{a\in\mathcal{A}_{k|k'}}w_{k|k'}^{a}\nonumber \\
 & \quad\times r_{k|k'}^{\pi_{a}^{(j)}(i),a^{\pi_{a}^{(j)}(i)}}p_{k|k'}^{\pi_{a}^{j}(i),a^{\pi_{a}^{(j)}(i)}}\left(x\right).\label{eq:PMB_PHD}
\end{align}
We can see both (\ref{eq:PMBM_PHD}) and (\ref{eq:PMB_PHD}) are independent
of the chosen permutations $\pi_{a}$ (or $\pi_{a}^{(j)}$), and are
therefore alike. Therefore, the PHD of the PMB does not change with
the iterations.

\subsection{Discussion\label{subsec:Discussion}}

It should be noted that the choice of $\pi$ does not affect the information
of the density (\ref{eq:MBM_permutation}). Nevertheless, it affects
the PMB approximation $\widetilde{q}\left(\cdot\right)$ that minimises
the KLD, see Lemma \ref{lem:Optimisation_q}. Therefore, the minimisation
over $\pi$ in (\ref{eq:KLD_optimisation_problem}), aims to find
$\pi$ that makes the PMB approximation $\widetilde{q}\left(\cdot\right)$
the most accurate, without losing information on the original density.

In \cite[Sec. III.C]{Williams15}, the V-PMB projection is obtained
by first introducing a missing distribution on the permutation of
the Bernoulli components for each global hypothesis. Then, leaving
out of consideration the PPP part, the KLD from the MBM posterior
to the MB approximation is upper bounded by using the log-sum inequality.
The resulting cost function depends on the cross-entropy for each
Bernoulli component and its approximation, the missing permutation
distributions, and a temperature $T\in[0,1]$. Then, the procedure
continues by minimising this KLD upper bound by coordinate descent,
iteratively optimising over the missing permutation distributions
and the MB approximation. In addition, reference \cite[Sec. III.D]{Williams15}
introduces an approximate, fast V-PMB projection by using a missing
distribution over the local hypotheses of each Bernoulli instead of
the missing distribution over the permutations, and then relaxing
the constraints of this missing distribution. 

In comparison, we have derived the V-PMB projection (not its approximation)
by considering an augmented space consisting of the global hypothesis
variable and the set of targets with track indices, and performing
a coordinate descent minimisation of the KLD from the PMBM to the
required PMB. The introduction of auxiliary variables using an augmented
space is widely-used in probabilistic inference, for instance, in
particle filtering \cite{Pitt99} and node opening in graph inference
\cite[Sec. 7.3]{Wymeersch_book07}. The auxiliary variable approach
has also been used in multi-target inference to derive the track-oriented
PMB filter by direct KLD minimisation \cite{Angel20_e}, and to obtain
PMB filters based on set-type belief propagation \cite{Kim24,Xia23}.
It should be noted that the (\ref{eq:optimisation_permutation}) and
(\ref{eq:convergence_criterion}) can be written in terms of cross-entropy
instead of KLD. In this case, it is required to add the units of the
hypervolume of the single-target state such that the cross-entropy
is well-defined \cite[App. A]{Kim24}.

\section{Gaussian implementation}

The PMB projection algorithm explained in Section \ref{sec:PMB-projection-KLD_minimisation}
can be applied to any PMBM/PMB filter. In this section, we explain
in more detail the Gaussian implementation, which is the standard
one for PMBM/PMB filters for point targets \cite{Williams15b,Williams15,Angel18_b}.
Since the PPP remains unchanged in the projection, we focus on the
projection of the multi-Bernoulli part. In the Gaussian implementation,
the single-target density of the $i$-th Bernoulli with local hypothesis
$a^{i}$ is Gaussian such that
\begin{align}
p_{k|k'}^{i,a^{i}}\left(x\right) & =\mathcal{N}\left(x;\overline{x}_{k|k'}^{i,a^{i}},P_{k|k'}^{i,a^{i}}\right)\label{eq:Gaussian_density_p}
\end{align}
where $\overline{x}_{k|k'}^{i,a^{i}}$ is its mean and $P_{k|k'}^{i,a^{i}}$
is its covariance matrix. The single-target density under permutation
$\pi_{a}$ is then
\begin{align}
p_{k|k'}^{\pi_{a}(i),a^{\pi_{a}(i)}}\left(x\right) & =\mathcal{N}\left(x;\overline{x}_{k|k'}^{\pi_{a}(i),a^{\pi_{a}(i)}},P_{k|k'}^{\pi_{a}(i),a^{\pi_{a}(i)}}\right).\label{eq:Gaussian_density_p_permuted}
\end{align}

In addition, we constrain the single-target densities of the projected
PMB density $\widetilde{q}^{(j)}\left(\cdot\right)$ to be Gaussian
of the form
\begin{align}
p^{(j),i}\left(x\right) & =\mathcal{N}\left(x;\overline{x}_{k|k'}^{(j),i},P_{k|k'}^{(j),i}\right).\label{eq:Gaussian_density_q}
\end{align}

\begin{lem}
\label{lem:Optimisation_q_Gaussian}Given a PMBM with Gaussian single-target
densities of the form (\ref{eq:Gaussian_density_p}) and the permutations
$\pi^{(j)}$ for all global hypotheses, the optimal PMB approximation
$\widetilde{q}^{(j)}\left(\cdot\right)$ of the form (\ref{eq:PMB_approx}),
with Gaussian single-target densities of the form (\ref{eq:Gaussian_density_q}),
in the sense (\ref{eq:optimisation_q}) is characterised by PPP intensity
$\lambda^{q}\left(\cdot\right)$ given by (\ref{eq:lambda_q}), probability
of existence $r^{(j),i}$ given by (\ref{eq:prob_existence_q}) and
\begin{align}
\overline{x}_{k|k'}^{(j),i} & =\frac{1}{r^{(j),i}}\sum_{a\in\mathcal{A}_{k|k'}}w_{k|k'}^{a}r_{k|k'}^{\pi_{a}^{(j)}(i),a^{\pi_{a}^{(j)}(i)}}\overline{x}_{k|k'}^{\pi_{a}(i),a^{\pi_{a}(i)}}\\
P_{k|k'}^{(j),i} & =\frac{1}{r^{(j),i}}\sum_{a\in\mathcal{A}_{k|k'}}w_{k|k'}^{a}r_{k|k'}^{\pi_{a}^{(j)}(i),a^{\pi_{a}^{(j)}(i)}}\left[P_{k|k'}^{\pi_{a}(i),a^{\pi_{a}(i)}}\vphantom{\left(\overline{x}_{k|k'}^{\pi_{a}(i),a^{\pi_{a}(i)}}-\overline{x}_{k|k'}^{(j),i}\right)^{T}}\right.\nonumber \\
 & \left.\quad+\left(\overline{x}_{k|k'}^{\pi_{a}(i),a^{\pi_{a}(i)}}-\overline{x}_{k|k'}^{(j),i}\right)\left(\overline{x}_{k|k'}^{\pi_{a}(i),a^{\pi_{a}(i)}}-\overline{x}_{k|k'}^{(j),i}\right)^{T}\right].
\end{align}
\end{lem}
This lemma is a direct result from Lemma \ref{lem:Optimisation_q}
and the KLD minimisation properties of moment matching for Gaussian
densities \cite{Bishop_book06}.

To optimise over the permutations, we first need to compute the KLDs
in (\ref{eq:optimisation_permutation}). The expressions of these
KLDs for the Gaussian implementation are provided in Appendix \ref{sec:AppendixC}.
Note that these KLDs are also required for the convergence criterion
in Section \ref{subsec:Convergence-criterion}. Then, we obtain the
optimal permutation for each global hypothesis following the procedure
provided in Algorithm \ref{alg:Optimisation_permutations}, where
$h^{i}$ denotes the number of local hypothesis of the $i$-th Bernoulli.
The pseudocode of the V-PMB projection is provided in Algorithm \ref{alg:VPMB_projection}.

\begin{algorithm}
\caption{\label{alg:Optimisation_permutations}Optimisation over the permutation
for each global hypothesis}

{\fontsize{9}{9}\selectfont

\textbf{Input: }PMBM $\widetilde{f}_{k|k'}\left(\cdot\right)$, see
(\ref{eq:PMBM_aux_var2}), and PMB $\widetilde{q}^{(j)}\left(\cdot\right)$,
see (\ref{eq:PMB_approx}), both with Gaussian single-target densities.

\textbf{Output: }Optimal permutations $\pi_{a}^{(j+1)}$ for all global
hypotheses $a\in\mathcal{A}_{k|k'}$, and cost $c^{(j)}$.

\begin{algorithmic}     

\For{$l=1$ to $n_{k|k'}$} \Comment{Go through Bernoullis in $\widetilde{q}^{(j)}$}

\For{$i=1$ to $n_{k|k'}$} \Comment{Go through Bernoullis in $\widetilde{f}_{k|k'}$}

\For{$a^i=1$ to $h^i$} \Comment{Go through all local hypotheses and compute KLD for the Bernoulli pair}

\State- $D(i,a^{i},l)=\mathrm{D}\left(f_{k|k'}^{i,a^{i}}\left\Vert q^{l}\right.\right)$,
see Appendix \ref{sec:AppendixC}. 

\EndFor

\EndFor

\EndFor

\For{$\forall a\in \mathcal{A}_{k|k'}$ }

\For{$l=1$ to $n_{k|k'}$}

\For{$i=1$ to $n_{k|k'}$}

\State- $C_{i,l}=D(i,a^{i},l)$. \Comment{Write element of cost matrix $C$.}

\EndFor

\EndFor

\State- Obtain $\pi_{a}^{(j+1)}$ and the cost $c_{a}^{(j)}$ by
solving the optimal 2-D assignment problem with cost matrix $C$ \cite{Crouse16}. 

\EndFor

\State- Compute cost $c^{(j)}$ using (\ref{eq:convergence_criterion}):
$c^{(j)}=\sum_{a\in\mathcal{A}_{k|k'}}w_{k|k'}^{a}c_{a}^{(j)}$.

\end{algorithmic}

}
\end{algorithm}

\begin{algorithm}
\caption{\label{alg:VPMB_projection}Variational PMB projection of a PMBM (Gaussian
implementation)}

{\fontsize{9}{9}\selectfont

\textbf{Input: }PMBM $\widetilde{f}_{k|k'}\left(\cdot\right)$, see
(\ref{eq:PMBM_aux_var2}), with Gaussian single-target densities,
convergence threshold $\Gamma$, maximum number of iterations $J$.

\textbf{Output: }PMB approximation $\widetilde{q}^{(j+1)}\left(\cdot\right)$. 

\begin{algorithmic}     

\State- Set $c^{(0)}=\infty$.

\State- Set $\pi_{a}^{(0)}=\left(1,...,n_{k|k'}\right)$ $\forall a$.

\State- Calculate PMB $\widetilde{q}^{(0)}\left(\cdot\right)$ using
$\pi_{a}^{(0)}$ in Lemma \ref{lem:Optimisation_q_Gaussian}.

\For{$j=0$ to $J-1$} 

\State- Calculate $\pi_{a}^{(j+1)}$ $\forall a$ and cost $c^{(j+1)}$
running Algorithm \ref{alg:Optimisation_permutations} with inputs
$\widetilde{f}_{k|k'}\left(\cdot\right)$ and $\widetilde{q}^{(j)}\left(\cdot\right)$.

\State- Calculate PMB $\widetilde{q}^{(j+1)}\left(\cdot\right)$
and $c^{(j)}$ using $\pi_{a}^{(j+1)}$ in Lemma \ref{lem:Optimisation_q_Gaussian}.

\State- If there is convergence, see (\ref{eq:convergence_criterion2}),
return $\widetilde{q}^{(j+1)}\left(\cdot\right)$.

\EndFor

\State- Return $\widetilde{q}^{(J)}\left(\cdot\right)$.

\end{algorithmic}

}
\end{algorithm}

\section{Simulations}

In this section, we compare the V-PMB filter\footnote{Code of the V-PMB filter is available at https://github.com/Agarciafernandez/MTT.},
with the PMBM filter and other PMB filter variants, specifically,
the GNN-PMB filter, and the (track-oriented) PMB filter, implemented
via Murty's algorithm \cite{Murty68} (M-PMB) and via belief propagation
(BP-PMB)\footnote{The BP-PMB implementation is provided in the ancillary files in https://arxiv.org/abs/1203.2995. }
\cite{Williams14}. The filters have been implemented with the following
parameters \cite{Angel18_b}: threshold for pruning the PPP weights
$\Gamma_{p}=10^{-5}$, threshold for pruning Bernoulli components
$\Gamma_{b}=10^{-5}$, estimator 1 with threshold 0.4, and ellipsoidal
gating with threshold 20. The PMBM filter considers a maximum number
of global hypotheses $N_{h}=200$. The PMBM, V-PMB and M-PMB filters
use Murty's algorithm \cite{Murty68} to select $N_{h}=200$ global
hypotheses in each update. In the V-PMB filter, to obtain the optimal
permutation in the iterated KLD optimisation, we use 2-D assignment
problem solver based on the modified Jonker-Volgenant algorithm provided
in the tracker component library \cite{Crouse17,Crouse16}. The convergence
threshold for the iterations is $\Gamma=0.1$. The BP-PMB implementation
uses a convergence threshold $10^{-4}$ and does not use gating. All
units are in the international system and are omitted for clarity.

\subsection{Models and scenario}

A target state is a vector with the target 2-D position and velocity
$[p_{x},v_{x},p_{y},v_{y}]^{T}$. The target moves with a nearly constant
velocity model with the single-target transition density
\begin{align}
g\left(x_{k}|x_{k-1}\right) & =\mathcal{N}\left(x_{k};Fx_{k-1},Q\right)
\end{align}
\begin{equation}
F=I_{2}\otimes\begin{bmatrix}1 & T\\
0 & 1
\end{bmatrix},\,Q=qI_{2}\otimes\begin{bmatrix}T^{3}/3 & T^{2}/2\\
T^{2}/2 & T
\end{bmatrix},
\end{equation}
where $\otimes$ is the Kronecker product, $q=0.01$, and the sampling
time $T=1$. The probability of survival is $p^{S}=0.99$. 

The sensor measures the positions of the targets. The density of a
single measurement given with the target state is
\begin{align}
l\left(z|x\right) & =\mathcal{N}\left(z;Hx,R\right)
\end{align}
\begin{equation}
H=I_{2}\otimes\begin{bmatrix}1 & 0\end{bmatrix},\,R=I_{2}.
\end{equation}
Clutter is uniformly distributed in the region of interest $A=[0,300]\times[0,300]$
with intensity $\lambda^{C}\left(z\right)=\overline{\lambda}^{C}\cdot u_{A}\left(z\right)$,
where $u_{A}\left(z\right)$ is a uniform density and $\overline{\lambda}^{C}=10$.
The probability of detection is $p^{D}=0.9$. All filters assume that
there are no targets at time 0. The PPP birth model intensity is Gaussian
with mean $\overline{x}_{k}^{b,1}=\left[100,0,100,0\right]^{T}$ and
covariance matrix $P_{k}^{b,1}=\mathrm{diag}\left(\left[150^{2},1,150^{2},1\right]\right)$,
with weight $w_{1}^{b,1}=3$ and $w_{k}^{b,1}=5\cdot10^{-3}$ for
$k>1$.

\begin{figure}
\begin{centering}
\includegraphics[scale=0.6]{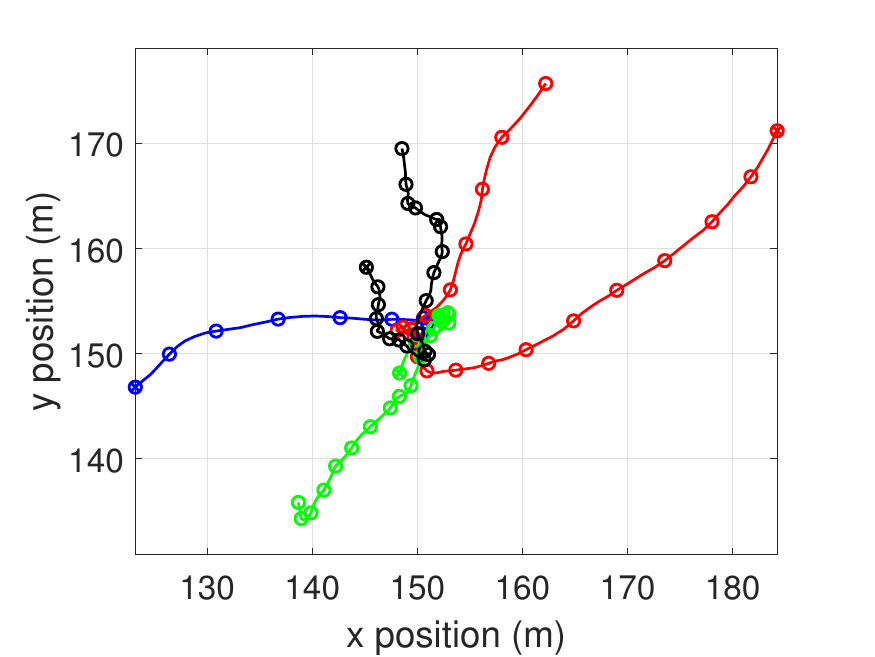}
\par\end{centering}
\caption{\label{fig:Scenario}Scenario of the simulations. The four targets
are born at the initial time step and remain alive for the entire
simulation, except for the blue target, which dies at time step 50
when all targets are near each other. Initial positions are marked
with crosses, and positions at 5-step intervals are marked with circles.}

\end{figure}

The scenario of the simulations is shown in Figure \ref{fig:Scenario}.
The scenario has 101 time steps and four targets that are in close
proximity in the middle of the simulation. The trajectories have been
generated running forward and backward dynamics from the middle of
the simulation, as in \cite[Sec. VI]{Williams15b}.

\subsection{Results}

Multi-target estimation performance is measured via the generalised
optimal sub-pattern assignment (GOSPA) metric to penalise the localisation
errors for detected targets, the number of missed targets, and the
number of false targets \cite{Rahmathullah17}. We set $p=2$, $c=10$,
and the Euclidean distance as base metric. We perform a Monte Carlo
simulation with 100 runs and calculate the root mean square GOSPA
(RMS-GOSPA) error at each time step. 

\begin{figure}
\begin{centering}
\includegraphics[scale=0.3]{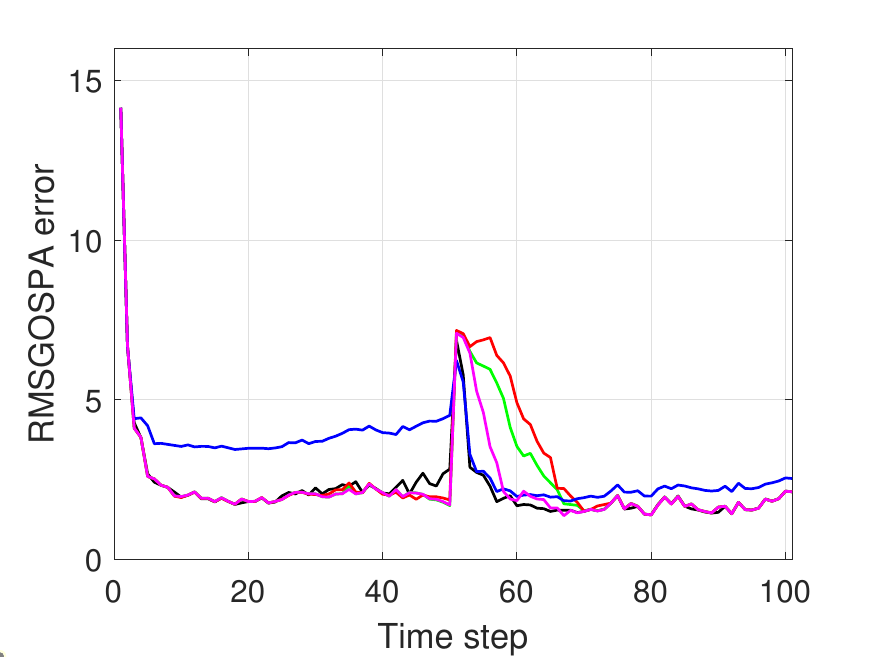}\includegraphics[scale=0.3]{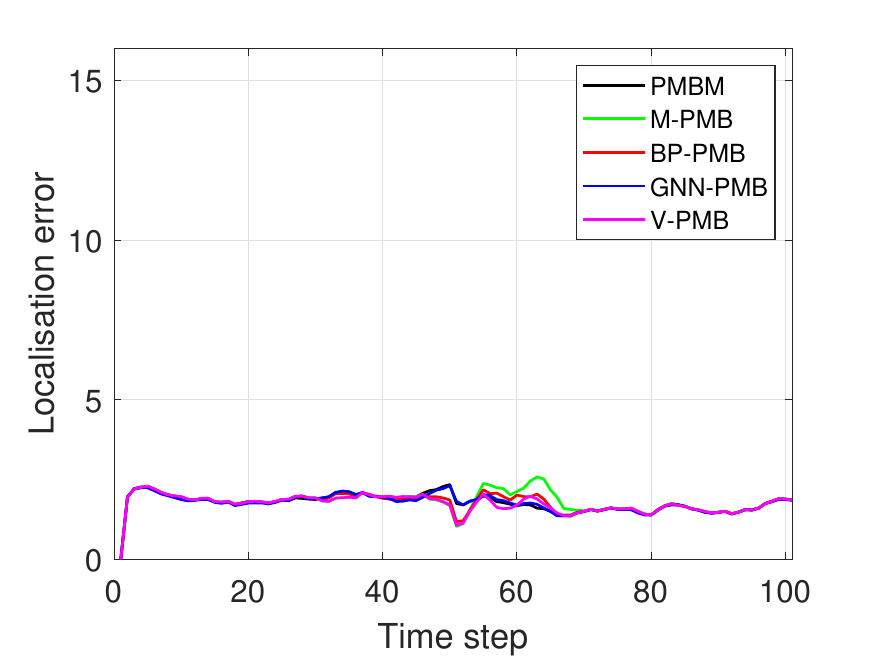}
\par\end{centering}
\begin{centering}
\includegraphics[scale=0.3]{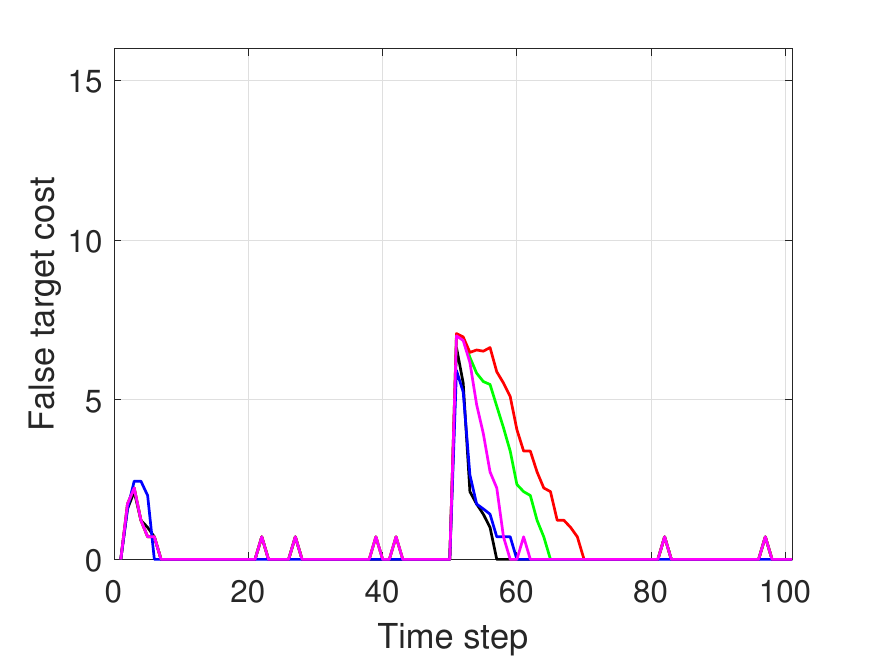}\includegraphics[scale=0.3]{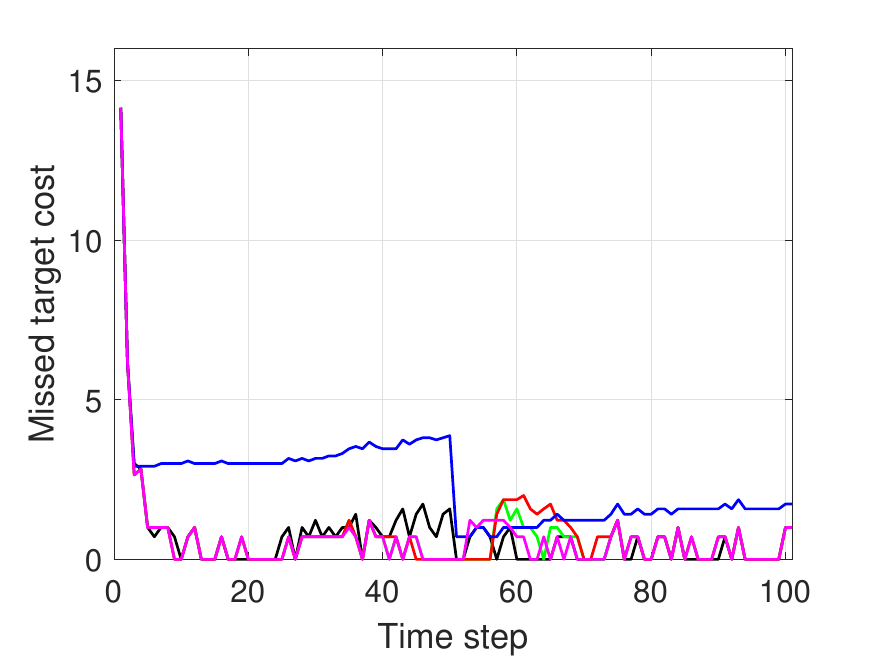}
\par\end{centering}
\caption{\label{fig:RMS-GOSPA-error}RMS-GOSPA error and its decomposition
at each time step. }

\end{figure}

The RMS-GOSPA error at each time step and its decomposition is shown
in Figure \ref{fig:RMS-GOSPA-error}. We can first see that the GNN-PMB
is the worst-performing filter in general, from the beginning of the
simulations. Since GNN-PMB only takes into account the most likely
data association hypothesis, this usually implies a loss in performance
compared to the other PMB filters. For the rest of the filters, the
main differences arise in the mid-point of the simulation, when targets
are in close proximity and one of them disappears. In fact, the main
differences in performance, as we can see from the GOSPA decomposition,
are in the false target cost, which basically measures how fast the
filters stop reporting detections from the target that has disappeared.
Overall, the PMBM filter is the best performing filter, which implies
that keeping 200 global hypotheses in the posterior at each time step
provides better results than the different PMB projections. The V-PMB
filter is the second best performing filter, so we can clearly see
the benefit of the iterations in the coordinate descent KLD optimisation.
Then, the second best PMB filter is the M-PMB filter, which outperforms
the BP-PMB filter. 

The average computational times (in seconds) to process all measurements,
obtained with a laptop with an Intel Core Ultra 7 155H processor,
are: 4.2 (PMBM), 0.4 (M-PMB), 0.3 (BP-PMB), 0.3 (GNN-PMB) and 1.9
(V-PMB). We can see PMBM is the algorithm that takes more time to
run, followed by V-PMB and M-PMB. BP-PMB and GNN-PMB are the fastest
implementations. 

We also show the RMS-GOSPA errors across all time steps for different
$p^{D}$ in Table \ref{tab:RMS-GOSPA-time}. Across all experiments,
PMBM is the best performing filter, and V-PMB is the best PMB filter
variant, followed by M-PMB, BP-PMB and GNN-PMB.

\begin{table}

\caption{\label{tab:RMS-GOSPA-time}RMS-GOSPA errors across all time steps}

\begin{centering}
\begin{tabular}{c|ccccc}
\hline 
 &
PMBM &
M-PMB &
BP-PMB &
GNN-PMB &
V-PMB\tabularnewline
\hline 
$p^{D}=0.9$ &
\uline{2.68} &
3.07 &
3.26 &
3.54 &
2.83\tabularnewline
$p^{D}=0.99$ &
\uline{2.34} &
2.66 &
2.85 &
2.83 &
2.46\tabularnewline
$p^{D}=0.8$ &
\uline{3.18} &
3.62 &
3.69 &
4.81 &
3.28\tabularnewline
$p^{D}=0.7$ &
\uline{3.66} &
4.03 &
4.10 &
5.82 &
3.67\tabularnewline
\hline 
\end{tabular}
\par\end{centering}
\end{table}

\section{Conclusions}

This paper has provided a new derivation of the V-PMB filter in \cite{Williams15,Williams14b},
valid for general detection-based measurement models \cite{Angel23}.
The derivation is based on writing the PMBM posterior in an augmented
space and then performing a coordinate descent KLD minimisation over
the permutations for each global hypothesis and PMB approximation.
We also show that the V-PMB projection keeps the PHD of the PMBM.
This new derivation provides a better understanding of the V-PMB projection
step and of the capabilities of using auxiliary variables for multi-target
probabilistic inference.

The V-PMB filter is beneficial compared to the track-oriented PMB
filters  when targets get in close proximity and then separate. This
has been demonstrated in a comparison with other PMB filter variants.
Nevertheless, the V-PMB filter requires higher computational complexity.
Future work can aim to speed-up the V-PMB filter, for instance, by
using approximate optimisation algorithms \cite{Williams15}, by clustering
Bernoulli components and applying the V-PMB projection in each cluster,
or by warm-starting the 2-D assignment solver \cite{Crouse16} from
one iteration to the next.

\appendices{}

\section{\label{sec:AppendixA}}

In this appendix, we prove Lemma \ref{lem:KLD_simplified}. The augmented
single target space $\mathbb{\mathbb{U}}_{k|k'}\times\mathbb{R}^{n_{x}}$
can be written as the disjoint union $\mathbb{\mathbb{U}}_{k|k'}\times\mathbb{R}^{n_{x}}=\uplus_{u=0}^{n_{k|k'}}\left\{ u\right\} \times\mathbb{R}^{n_{x}}$.
Thus, given a set $\widetilde{X}_{k}\subset\mathbb{\mathbb{U}}_{k|k'}\times\mathbb{R}^{n_{x}}$,
we can write $\widetilde{X}_{k}=\widetilde{Y}_{k}\uplus\widetilde{X}_{k}^{1}\uplus...\uplus\widetilde{X}_{k}^{n_{k|k'}}$,
where $\widetilde{Y}_{k}\subset\left\{ 0\right\} \times\mathbb{R}^{n_{x}}$
and $\widetilde{X}_{k}^{i}\subset\left\{ i\right\} \times\mathbb{R}^{n_{x}}$.
Then, the KLD (\ref{eq:KLD_augmented_space}) can be written as multiple
set integrals over these disjoint spaces \cite[Eq. (3.53)]{Mahler_book14}
\begin{align}
 & \mathrm{D}\left(\widetilde{f}_{k|k'}^{\pi}\left\Vert \widetilde{q}\right.\right)\nonumber \\
 & =\sum_{a\in\mathcal{A}_{k|k'}}\int\int...\int\widetilde{f}_{k|k'}^{\pi}\left(\widetilde{Y}_{k}\uplus\widetilde{X}_{k}^{1}\uplus...\uplus\widetilde{X}_{k}^{n_{k|k'}},a\right)\nonumber \\
 & \times\log\frac{\widetilde{f}_{k|k'}^{\pi}\left(\widetilde{Y}_{k}\uplus\widetilde{X}_{k}^{1}\uplus...\uplus\widetilde{X}_{k}^{n_{k|k'}},a\right)}{\widetilde{q}\left(\widetilde{Y}_{k}\uplus\widetilde{X}_{k}^{1}\uplus...\uplus\widetilde{X}_{k}^{n_{k|k'}},a\right)}\delta\widetilde{Y}_{k}\delta\widetilde{X}_{k}^{1}...\delta\widetilde{X}_{k}^{n_{k|k'}}\\
 & =\sum_{a\in\mathcal{A}_{k|k'}}w_{k|k'}^{a}\int\int...\int\widetilde{f}_{k|k'}^{p}\left(\widetilde{Y}_{k}\right)\nonumber \\
 & \times\prod_{i=1}^{n_{k|k'}}\left[\widetilde{f}_{k|k'}^{\pi_{a}(i),a^{\pi_{a}(i)}}\left(\widetilde{X}_{k}^{i}\right)\right]\nonumber \\
 & \times\log\frac{\widetilde{f}_{k|k'}^{p}\left(\widetilde{Y}_{k}\right)w_{k|k'}^{a}\prod_{i=1}^{n_{k|k'}}\left[\widetilde{f}_{k|k'}^{\pi_{a}(i),a^{\pi_{a}(i)}}\left(\widetilde{X}_{k}^{i}\right)\right]}{w_{q}^{a}\widetilde{q}^{p}\left(\widetilde{Y}_{k}\right)\prod_{i=1}^{n_{k|k'}}\left[\widetilde{q}^{i}\left(\widetilde{X}_{k}^{i}\right)\right]}\nonumber \\
 & \times\delta\widetilde{Y}_{k}\delta\widetilde{X}_{k}^{1}...\delta\widetilde{X}_{k}^{n_{k|k'}}\\
 & =\mathrm{D}\left(\widetilde{f}_{k|k'}^{p}\left\Vert \widetilde{q}^{p}\right.\right)+\sum_{a\in\mathcal{A}_{k|k'}}w_{k|k'}^{a}\log\frac{w_{k|k'}^{a}}{w_{q}^{a}}\nonumber \\
 & +\sum_{a\in\mathcal{A}_{k|k'}}w_{k|k'}^{a}\int\int...\int\prod_{i=1}^{n_{k|k'}}\left[\widetilde{f}_{k|k'}^{\pi_{a}(i),a^{\pi_{a}(i)}}\left(\widetilde{X}_{k}^{i}\right)\right]\nonumber \\
 & \,\times\log\frac{\prod_{i=1}^{n_{k|k'}}\left[\widetilde{f}_{k|k'}^{\pi_{a}(i),a^{\pi_{a}(i)}}\left(\widetilde{X}_{k}^{i}\right)\right]}{\prod_{i=1}^{n_{k|k'}}\left[\widetilde{q}^{i}\left(\widetilde{X}_{k}^{i}\right)\right]}\delta\widetilde{X}_{k}^{1}...\delta\widetilde{X}_{k}^{n_{k|k'}}\\
 & =\mathrm{D}\left(\widetilde{f}_{k|k'}^{p}\left\Vert \widetilde{q}^{p}\right.\right)+\mathrm{D}\left(w_{k|k'}^{a}\left\Vert w_{q}^{a}\right.\right)\nonumber \\
 & \quad+\sum_{a\in\mathcal{A}_{k|k'}}w_{k|k'}^{a}\sum_{i=1}^{n_{k|k'}}\mathrm{D}\left(\widetilde{f}_{k|k'}^{\pi_{a}(i),a^{\pi_{a}(i)}}\left\Vert \widetilde{q}^{i}\right.\right).
\end{align}
This finishes the proof of Lemma \ref{lem:KLD_simplified}.

\section{\label{sec:AppendixB}}

In this appendix, we prove Lemma \ref{lem:Optimisation_q}. As the
KLD can be written as in (\ref{eq:KLD_simplified}), it is direct
to notice that $\widetilde{q}^{p}\left(\cdot\right)=\widetilde{f}_{k|k'}^{p}\left(\cdot\right)$.
In addition, as the KLD is additive for the Bernoulli components,
we can optimise for each Bernoulli component independently. If we
focus on the optimisation over the $i$-th Bernoulli component, we
can write the KLD as
\begin{align}
\mathrm{D}\left(\widetilde{f}_{k|k'}^{\pi^{(j)}}\left\Vert \widetilde{q}\right.\right) & =\mathrm{z}-\sum_{a\in\mathcal{A}_{k|k'}}w_{k|k'}^{a}\int\left[\widetilde{f}_{k|k'}^{\pi_{a}^{(j)}(i),a^{\pi_{a}^{(j)}(i)}}\left(\widetilde{X}_{k}^{i}\right)\right]\nonumber \\
 & \quad\times\log\widetilde{q}^{i}\left(\widetilde{X}_{k}^{i}\right)\delta\widetilde{X}_{k}^{i}
\end{align}
where $\mathrm{z}$ is a constant that does not depend on $\widetilde{q}^{i}\left(\cdot\right)$.

By standard KLD minimisation, the density $\widetilde{q}^{i}\left(\cdot\right)$
that minimises this KLD is
\begin{align}
\widetilde{q}^{i}\left(\widetilde{X}_{k}^{i}\right) & =\sum_{a\in\mathcal{A}_{k|k'}}w_{k|k'}^{a}\left[\widetilde{f}_{k|k'}^{\pi_{a}^{(j)}(i),a^{\pi_{a}^{(j)}(i)}}\left(\widetilde{X}_{k}^{i}\right)\right].\label{eq:q_optim_appendix}
\end{align}
Plugging (\ref{eq:Bernoulli_density_filter-aux}) into (\ref{eq:q_optim_appendix})
yields the probability of existence in (\ref{eq:prob_existence_q})
and the single-target density in (\ref{eq:single_target_density_q})
proving Lemma \ref{lem:Optimisation_q}.

\section{\label{sec:AppendixC}}

In this section, for completeness, we give the expression of the KLD
$\mathrm{D}\left(\widetilde{f}_{k|k'}^{\pi_{a}(i),a^{\pi_{a}(i)}}\left\Vert \widetilde{q}^{(j),i}\right.\right)$.
To do this, we first provide the KLD between Gaussian single-target
densities (\ref{eq:Gaussian_density_p_permuted}) and (\ref{eq:Gaussian_density_q})
which is given by \cite{Hershey07}
\begin{align}
 & \mathrm{D}\left(p_{k|k'}^{\pi_{a}(i),a^{\pi_{a}(i)}}\left\Vert p^{(j),i}\right.\right)=\left[\mathrm{tr}\left(\left(P_{k|k'}^{(j),i}\right)^{-1}P_{k|k'}^{\pi_{a}(i),a^{\pi_{a}(i)}}\right)\right.\nonumber \\
 & -\log\left(\frac{\left|P_{k|k'}^{\pi_{a}(i),a^{\pi_{a}(i)}}\right|}{\left|P_{k|k'}^{(j),i}\right|}\right)-n_{x}+\left(\overline{x}_{k|k'}^{(j),i}-\overline{x}_{k|k'}^{\pi_{a}(i),a^{\pi_{a}(i)}}\right)^{T}\nonumber \\
 & \left.\times\left(P_{k|k'}^{(j),i}\right)^{-1}\left(\overline{x}_{k|k'}^{(j),i}-\overline{x}_{k|k'}^{\pi_{a}(i),a^{\pi_{a}(i)}}\right)\right]/2.
\end{align}
Then, the KLD between Bernoulli densities, for $r^{(j),i}\not\in\{0,1\}$,
is \cite{Fontana23}
\begin{align}
 & \mathrm{D}\left(\widetilde{f}_{k|k'}^{\pi_{a}(i),a^{\pi_{a}(i)}}\left\Vert \widetilde{q}^{(j),i}\right.\right)\nonumber \\
 & =\left(1-r_{k|k'}^{\pi_{a}^{(j)}(i),a^{\pi_{a}^{(j)}(i)}}\right)\log\frac{1-r_{k|k'}^{\pi_{a}^{(j)}(i),a^{\pi_{a}^{(j)}(i)}}}{1-r^{(j),i}}\nonumber \\
 & \quad+r_{k|k'}^{\pi_{a}^{(j)}(i),a^{\pi_{a}^{(j)}(i)}}\log\frac{r_{k|k'}^{\pi_{a}^{(j)}(i),a^{\pi_{a}^{(j)}(i)}}}{r^{(j),i}}\nonumber \\
 & \quad+r_{k|k'}^{\pi_{a}^{(j)}(i),a^{\pi_{a}^{(j)}(i)}}\mathrm{D}\left(p_{k|k'}^{\pi_{a}(i),a^{\pi_{a}(i)}}\left\Vert p^{(j),i}\right.\right),
\end{align}
and, for $r_{k|k'}^{\pi_{a}^{(j)}(i),a^{\pi_{a}^{(j)}(i)}}=r^{(j),i}\in\{0,1\},$
is
\begin{align}
\mathrm{D}\left(\widetilde{f}_{k|k'}^{\pi_{a}(i),a^{\pi_{a}(i)}}\left\Vert \widetilde{q}^{i}\right.\right) & =r_{k|k'}^{\pi_{a}^{(j)}(i),a^{\pi_{a}^{(j)}(i)}}\mathrm{D}\left(p_{k|k'}^{\pi_{a}(i),a^{\pi_{a}(i)}}\left\Vert p^{(j),i}\right.\right).
\end{align}

\bibliographystyle{IEEEtran}
\bibliography{6C__Trabajo_laptop_Referencias_Referencias}

\end{document}